\documentclass[aps,prb,twocolumn,superscriptaddress]{revtex4}
\usepackage{ae}
\usepackage[T1]{fontenc}
\usepackage[ansinew]{inputenc}
\usepackage{amsmath}
\usepackage{amssymb}
\usepackage{graphicx}
\usepackage{color}

\usepackage[colorlinks]{hyperref}
\hypersetup{colorlinks=true,urlcolor=blue}

\usepackage{epstopdf}

\def\sig{{\mbox{\boldmath{$\sigma$}}}}

\def\el{{\mbox{\boldmath{$\ell$}}}}

\def\sig{{\mbox{\boldmath{$\sigma$}}}}
\usepackage{soul,xcolor}
\setstcolor{red}                   % Change color of strike-through line created by "\st" here.

\def\sig{{\mbox{\boldmath{$\sigma$}}}}

\begin{document}

\date{\today}

\title{
Magnetization generated by microwave-induced Rashba interaction}

%\author{{\color{blue}{\bf The order of names is open for changes,   }}}

\author{O. Entin-Wohlman}
\email{orawohlman@gmail.com}
\affiliation{School of Physics and Astronomy, Tel Aviv University, Tel Aviv 69978, Israel}
%\affiliation{Physics Department, Ben Gurion University, Beer Sheva 84105, Israel}

\author{R. I. Shekhter}
\affiliation{Department of Physics, University of Gothenburg, SE-412
96 G{\" o}teborg, Sweden}

\author{M. Jonson}
\affiliation{Department of Physics, University of Gothenburg, SE-412
96 G{\" o}teborg, Sweden}

\author{A. Aharony}
\affiliation{School of Physics and Astronomy, Tel Aviv University, Tel Aviv 69978, Israel}

%%%%%%%%%%%%%%%%%%%%%%%%%%%%%%%%%%%%%%%%%%%%%%%%%%%%%%
%%%%%%%%%%%%%%%%%%%%%%%%%%%%%%%%%%%%%%%%%%%%%%%%%%%%%%

\begin{abstract}

We show that a controllable dc magnetization  is accumulated in a junction comprising a quantum dot coupled to non-magnetic reservoirs if the junction is subjected to a time-dependent spin-orbit interaction. The latter is induced by an ac electric field generated by microwave irradiation of the gated junction.
The magnetization is caused by inelastic spin-flip scattering of electrons that tunnel through the junction, and depends on the polarization of the electric field: a circularly polarized field leads to the maximal effect, while there is no effect in a linearly polarized  field.
 Furthermore, the magnetization increases as a step function (smoothened by temperature) as the microwave photon energy becomes larger than the absolute value of the difference between the single energy level on the quantum dot and the common chemical potential in the leads.

\end{abstract}

\maketitle

%%%%%%%%%%%%%%%%%%%%%%%%%%%%%%%%%%%%%%%%%%%%%%%%%%%%%%
%%%%%%%%%%%%%%%%%%%%%%%%%%%%%%%%%%%%%%%%%%%%%%%%%%%%%%

\section{Introduction}

%%%%%%%%%%%%%%%%%%%%%%%%%%%%%%%%%%%%%%%%%%%%%%%%%%%%%%
%%%%%%%%%%%%%%%%%%%%%%%%%%%%%%%%%%%%%%%%%%%%%%%%%%%%%%

The possibility to create and manipulate magnetic order confined  to the nanometer length scale is currently attracting interest  because  of possible implications for magnetic devices and material developments \cite{Sander}.
Such a confined magnetization is seldom
achieved   by applying an external magnetic field, due to practical difficulties encountered when attempting  to spatially localize the field.
It can,  however, be realized
by modulating  the exchange-interaction strength, for instance
along a depth-profile variation of certain alloys' constituents \cite{Kirby}.
In contrast to external magnetic fields, electrical currents can be localized quite easily when  injected from
nanometer-size electric weak links (e.g.,  quantum point contacts).
In case such currents are spin polarized, as happens for electrons  injected from magnetic materials,  they lead to the creation of magnetic torques that can be exploited to manipulate  and control the local magnetization of a ferromagnet \cite{Miron}. Spin injection of ac and dc currents from ferromagnetic materials were indeed  detected and imaged \cite{Pile}.
Yet another tool for efficient manipulation of magnetic order in nano-scale devices depends on the  interplay between charge and spin brought about by the spin-orbit interaction \cite{Araki} which couples  the spin and the momentum of the electrons. This is the so-called ``spin-charge conversion" or the Edelstein-Rashba effect \cite{Rashba,Edelstein}, which occurs at interfaces where the Rashba spin-orbit interaction is active \cite{Rojas,Salemi}.

%%%%%%%%%%%%%%%%%%%%%%%%%%%%%%%%%%%%%%%%%%%%%%%%%%%%%%
%%%%%%%%%%%%%%%%%%%%%%%%%%%%%%%%%%%%%%%%%%%%%%%%%%%%%%

The phenomenon of spin-charge conversion at an interface with broken inversion-symmetry
has also been  achieved by shining light on the sample \cite{Puebla,Hernangomez}. In these configurations the radiated field couples equally  to both spin components, and the spin selectivity needed for the spin-charge conversion is procured  by the presence of a (static) Rashba interaction at the irradiated interface.
We propose in this paper a different scenario:
 the possibility to magnetize  initially {\em spin-inactive}   conducting nanostructures through
 a Rashba interaction {\em induced} by
 an ac electric field
 generated by external microwave radiation.
 Put differently, the generated electric field couples the momenta of the electrons with their spins.
 Employing  an ac electric field to induce the Rashba interaction on  nanostructures modifies qualitatively and profoundly the electrons' kinematics in them. The inelastic transitions of electrons that tunnel  through the junction acquire a spin dependence due to a correlation between photon absorption and emission processes and distinct spin-flip transitions. This paves a way to magnetize a spin-inactive material in the absence of external magnetic fields.

%%%%%%%%%%%%%%%%%%%%%%%%%%%%%%%%%%%%%%%%%%%%%%%%%%%%%%
%%%%%%%%%%%%%%%%%%%%%%%%%%%%%%%%%%%%%%%%%%%%%%%%%%%%%%

Once the   Rashba interaction is established in the junction,   the tunneling amplitudes are augmented  by the  Aharonov-Casher \cite{AC} phase factors
which in turn render the tunneling to be accompanied by spin flips \cite{PRL2013}. Namely, the Aharonov-Casher factors can be considered as unitary rotations of the magnetic moment.
This by itself is insufficient to produce spin selectivity, as follows from considerations based on time-reversal symmetry \cite{Bardarson}.
However, the   ac electric field generates a  Rashba interaction which depends on time, thus breaks time-reversal symmetry
 and makes  spin-selective tunneling possible. We have recently observed that such  time-dependent tunneling   can result in the appearance of a dc electromotive force on the junction \cite{R2019}.
In this paper we show that   spin-selective  transport between non-magnetic conductors is created when the Rashba interaction is induced by an oscillating electric field, and leads to the accumulation of a dc magnetic order, even when the junction is unbiased.  The magnitude of the induced magnetization depends on  the polarization of the electric field, and reaches its maximal value for a  circularly polarized field.  Accordingly,    a totally non-magnetic conductor can be magnetized when subjected to a rotating electric field.

The paper is divided into two parts. We first analyse in Sec. \ref{AMs}
the simplest possible junction, which comprises
a quantum dot coupled to a single metal reservoir, as shown in Fig. \ref{single_lead}.
We derive there the dc magnetization on the dot and the  rate by which  a  magnetic order is built up in the lead. %Given these two quantities,  one wonders about  a possible relationship between the two.
The total magnetization in the junction is not expected to be conserved   when a time-dependent Rashba interaction is active. However,
when an electron moves via the spin-orbit-active link from the dot to the reservoir, its magnetization rotates by the Aharonov-Casher factor to a new direction. Therefore (as we show in Sec. \ref{AMs}),  the sum of  the time-derivatives of the magnetization in the dot along
an arbitrary direction $\hat{\el}$, and that of the magnetization in the lead along the direction $\hat{\el}''_L(t)$, obtained from $\hat{\el}$ after  rotating it by the Aharonov-Casher factors, is zero, namely the two magnetization rates
cancel one another.

In the second part of the paper, Sec. \ref{2r},  we consider  a configuration where the dot is coupled to two reservoirs, see Fig. \ref{sys}. These  can be kept at different chemical potentials (or temperatures), which provides another tool for controlling the system.
Not surprisingly (in view of the results in Sec. \ref{AMs}), the magnetization accumulated on the dot in this case  depends on electron tunneling  from both leads. It hinges on the chemical potential and temperature of each lead via the Fermi distribution there. Note, though, that its existence does not necessitate a chemical potential difference, or a temperature difference, between the two leads.
The dc rate of change of the magnetization in each of  the leads, however, is modified qualitatively as compared to the one found  in  Sec. \ref{AMs} for a dot connected to a single lead:  a voltage bias across the junction, or a temperature difference between the two leads,  allows for an `extra' dc magnetization in one lead, at the expense of the other lead.  Similar to the findings in Sec. \ref{AMs}, the total magnetization in the system is not conserved,
but the magnetization rates  along appropriate rotated directions can add up to zero.

Technical details of the calculation are
relegated to the Appendix. There, calculations are carried out  for the second configuration, depicted in Fig. \ref{sys}, since it is  straightforward to infer from those the relations needed for the first configuration, depicted in Fig. \ref{single_lead}. For this reason, our notations in Sec. \ref{AMs} assign the letter $L$ to the physical characteristics of the single lead.

%%%%%%%%%%%%%%%%%%%%%%%%%%%%%%%%%%%%%%%%%%%%%%%%%%%%%%%
%%%%%%%%%%%%%%%%%%%%%%%%%%%%%%%%%%%%%%%%%%%%%%%%%%%%%%

\section{spin in a single-lead junction}
\label{AMs}

\begin{figure}[htp]
\includegraphics[width=8cm]{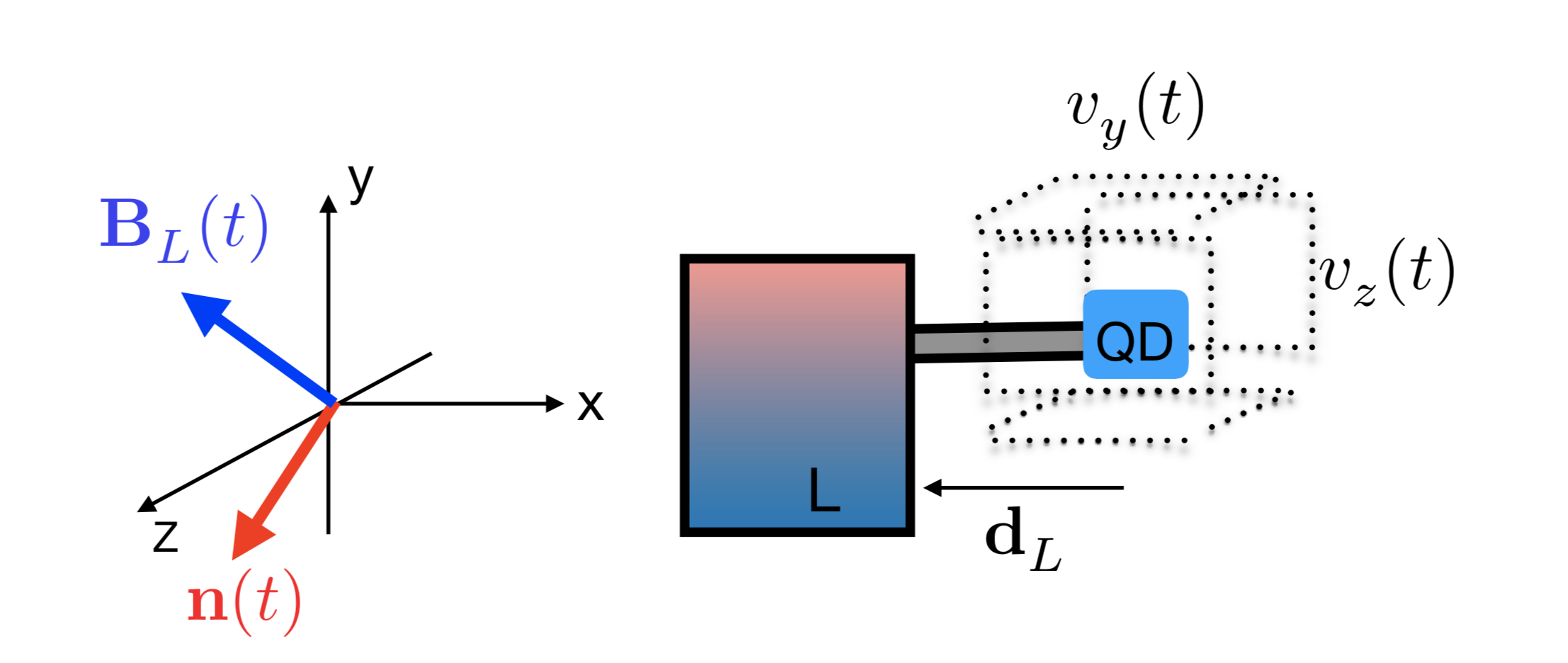}
\caption{(Color online.)    A quantum dot, represented by a single localized energy level, is attached to a non-magnetic metal lead by  a weak link  along  the ${\bf x}$-axis. The four plates represent  the application of microwave-induced ac gate voltages, $v_{y}(t)$ and $v_{z}(t)$, which create time-dependent electric fields  along the $\hat{\bf y}$ and $\hat{\bf z}$ directions, respectively.  The resulting total  electric field  along the vector $\hat{\bf n}(t)$ can be made to  rotate in the $y$-$z$ plane by introducing a phase shift between the oscillating gate voltages.  The electric field induces  a  Rashba interaction in the weak link, that is represented by the effective magnetic field ${\bf B}^{}_{L}(t)$, which is perpendicular to both $\hat{\bf x}$ and $\hat{\bf n}(t)$.  }
\label{single_lead}
\end{figure}

We begin by considering a  quantum dot coupled to just a single,  non-magnetic,  metal lead by a weak link, as depicted in Fig. \ref{single_lead}. This,  the simplest  configuration of interest here,  serves  to demonstrate the building up of a magnetic moment in the dot and in the lead under the effect of a  rotating electric field.

By applying  microwave-induced time-dependent gate voltages
as indicated in Fig. \ref{single_lead},    an ac   electric field is exerted on the weak link. The field is oriented along the vector $\hat{\bf n}(t)$, which  rotates with the microwave  frequency $\Omega$ in the $y$-$z$ plane,
\begin{align}
\hat{\bf n}(t)=\hat{\bf z}\cos(\Omega t)-\gamma\hat{\bf y} \sin(\Omega t)\ .
\end{align}
Here,  $\gamma$ is the parameter that measures the deviation from  perfectly circular polarization: for $\gamma=1$ (or  $\gamma=-1$) the electric field is  circularly polarized, rotating in a clockwise (or anti-clockwise) direction with respect to the positive ${\bf x}$-direction. For $\gamma=0$  the field is  linearly polarized. The significance of $\gamma$ is elucidated below.

%%%%%%%%%%%%%%%%%%%%%%%%%%%%%%%%%%%%%%%%%%%%%%%%%%%%%%%%%%%%
%%%%%%%%%%%%%%%%%%%%%%%%%%%%%%%%%%%%%%%%%%%%%%%%%%%%%%%%%%%%

In a weak link with broken inversion symmetry \cite{Rashba}, the electric field creates a time-dependent Rashba interaction \cite{Duckheim}, which manifests itself  in the form of a phase factor superimposed on the tunneling amplitude. This phase factor, arising from the Aharonov-Casher effect \cite{AC}, reads
\begin{align}
V^{}_{L}(t)&=\exp[ik^{}_{\rm so}{\bf d}^{}_{L}\times\hat{\bf n}(t)\cdot\sig]\ ,
\label{acpf}
\end{align}
where ${\bf d}^{}_{L}=-d^{}_{L}\hat{\bf x}
$
 is the radius-vector from the  dot to the  lead, see Fig. \ref{single_lead}.
In Eq. (\ref{acpf}), $\sig=[\sigma^{}_{x},\sigma^{}_{y},\sigma^{}_{z}]$ is the vector of the Pauli matrices, and $k_{\rm so}^{}$ represents the strength of the Rashba spin-orbit interaction  (in inverse-length units), which is proportional to the ac electric field associated with the microwave radiation. 
The tunneling Hamiltonian that describes transitions between
electronic states in the  lead (given by the  operator $c^{\dagger}_{{\bf k}\sigma}$ that
creates  an electron of energy $\epsilon^{}_{k}$, wave vector ${\bf k}$,   and spin index $\sigma$) and those  on the  dot
(given by the operator $d^{\dagger}_{\sigma'}$ that creates an electron of energy $\epsilon$ with spin index $\sigma'$) is
\begin{align}
{\cal H}^{L}_{\rm tun}(t)&=J^{}_{L}\sum_{\sigma,\sigma'}[V^{\ast}_{L}(t)]^{}_{\sigma\sigma'}\sum_{\bf k}d^{\dagger}_{\sigma'}c^{}_{{\bf k}\sigma}+{\rm H.c.}\nonumber\\
&\sim J^{}_{L}\sum_{\sigma,\sigma'}\Big ([1-|{\bf B}^{}_{L}(t)|^{2}/2]\delta^{}_{\sigma,\sigma'}\nonumber\\
&-i[\sig\cdot{\bf B}^{}_{L}(t)]^{}_{\sigma'\sigma}\Big )\sum_{\bf k}d^{\dagger}_{\sigma'}c^{}_{{\bf k}\sigma}+{\rm H.c.}\ ,
\label{HtunL}
\end{align}
up to second order in the spin-orbit coupling $\alpha^{}_{L}=k^{}_{\rm so}d^{}_{L}$ ($J^{}_{L}$ is the tunneling energy scale). The spin-orbit interaction appears as   a dimensionless  effective magnetic field oscillating with frequency $\Omega$,
\begin{align}
{\bf B}^{}_{L}(t)
&=e^{i\Omega t}{\bf B}^{-}_{L}
+
e^{-i\Omega t}{\bf B}^{+}_{L}
\ ,\nonumber\\
{\bf B}^{\pm}_{L}&=(\alpha^{}_{L}/2)[\hat{\bf y}\pm i \gamma\hat{\bf z}]\ ,
\label{BLt}
\end{align}
that is perpendicular to the direction of the weak link, see Fig. \ref{single_lead}.

To this order in $\alpha^{}_L$, one identifies two processes  in Eq. (\ref{HtunL}). The first conserves the electronic spin during   tunneling,  while the second, the effective Zeeman term, involves spin flips accompanied by the absorption or emission of an energy quantum  $\Omega$ from the electric field \cite{com0}, as manifested in Eqs. (\ref{BLt}),
 using  $\hbar=1$.
At very low temperatures  the absorption transitions  dominate (for both the $d^{\dagger}_{\sigma'}c^{}_{{\bf k}\sigma}$ term and its hermitian conjugate)   in which case Eq. (\ref{HtunL}) simplifies. In particular,
$$
\sig\cdot{\bf B}^{}_{L}(t) \approx e^{-i\Omega t}\sig\cdot{\bf B}^{+}_{L} = e^{-i\Omega t} (\alpha^{}_{L}/2)(\sigma^{}_y + i \gamma \sigma^{}_z) \ .
$$
Note that
$$
\sigma^{}_y + i \gamma \sigma^{}_z = \sigma^+ (1+\gamma)/2 + \sigma^- (1-\gamma)/2\ ,
$$
where $\sigma^\pm = \sigma_y \pm i \sigma_z$ are operators that increase ($+$) and lower ($-$) the spin projection in the $\hat{\bf x}$-direction.
We may now infer that in a circularly polarized electric field, rotating in the clockwise direction ($\gamma=+1$),    absorption transitions lead to an accumulation on the dot of spins whose projections on the $\hat{\bf x}$-axis are positive (spin up), while if the electric field rotates in the anti-clockwise direction ($\gamma = -1$) absorption transitions lead to an accumulation of spins whose projections on the $\hat{\bf x}$-axis are negative (spin down). In a linearly polarized field ($\gamma = 0$) there is no preference for either spin projection and no net spin is accumulated.  Obviously these qualitative arguments will have to be verified by a detailed calculation, which is carried out  in the following.

Quite generally, the magnetization on the dot, given by the (dimensionless) vector ${\bf M}_{d}(t)$ (in units of $-g\mu^{}_{\rm B}/2$, where $g$ is the g-factor of the electron and $\mu_{\rm B}$ is the Bohr magneton), is {\it a priori} time-dependent,
 \begin{align}
 {\bf M}^{}_{d}(t)=\sum_{\sigma,\sigma '}\langle d^{\dagger}_{\sigma}(t)[\sig]^{}_{\sigma\sigma'}d^{}_{\sigma'}(t)\rangle\ ,
 \label{Md}
 \end{align}
and the angular brackets denote quantum averaging
with respect to  the
  Hamiltonian of the  junction,
\begin{align}
{\cal H}(t)={\cal H}^{}_{0}+{\cal H}^{L}_{\rm tun}(t)\ .
\label{HOm}
\end{align}
The time-independent Hamiltonian ${\cal H}^{}_{0}$ pertains to the decoupled system,
\begin{align}
{\cal H}^{}_{0}=
\sum_{\sigma}\epsilon d^{\dagger}_{\sigma}d^{}_{\sigma}+\sum_{{\bf k} ,\sigma}\epsilon^{}_{ k}c^{\dagger}_{{\bf k}\sigma}c^{}_{{\bf k}\sigma}\ ,
\label{H0m}
\end{align}
with the first term  describing the decoupled dot
and the second  the decoupled electronic reservoir,  assumed to consist of non-polarized free electrons;
${\cal H}^{L}_{\rm tun}(t)$ is given in Eq. (\ref{HtunL}).
 The quantum average in Eq. (\ref{Md}) is  related to  the lesser Keldysh Green's function on the dot at equal times, defined as
 %\begin{align}
% \langle d^{\dagger}_{\sigma}(t)d^{}_{\sigma'}(t)\rangle
% \equiv -i[G^{<}_{dd,L}(t,t)]^{}_{\sigma'\sigma}\ .
% \label{sd}
% \end{align}
 %\mats{[MJ: This is an implicit definition of the lesser Green's function since it appears not by itself and on the RHS of the equivalence sign. How about,
 \begin{align}
[G^{<}_{dd,L}(t,t)]^{}_{\sigma'\sigma}  \equiv i  \langle d^{\dagger}_{\sigma}(t)d^{}_{\sigma'}(t)\rangle
 \label{sd}
 \end{align}
 %The order of the spin indices are reversed between LHS and RHS. Is that correct?}
 This Green's function is derived \cite{com1} in Appendix \ref{Technical}, exploiting the Keldysh technique. %\cite{Langreth,Jauho}.
 Inserting  Eq. (\ref{Gd2s}) into the definition  (\ref{Md}), one finds
  \begin{align}
 {\bf M}^{}_{d}(t)=2\Gamma^{}_{L}\int\frac{d\omega}{2\pi}f^{}_{L}(\omega){\rm Tr}\{W^{}_{L}(t,\omega)\sig\}\ ,
 \label{Mdtr}
 \end{align}
where the trace is carried out in spin space. Here,  $\Gamma_{L}$ is the width of the Breit-Wigner
 resonance formed on the dot due to the coupling with the lead \cite{com1}, and $f^{}_{L}(\omega)$ is the Fermi distribution in the lead.

\begin{figure}[htp]
\includegraphics[width=7cm]{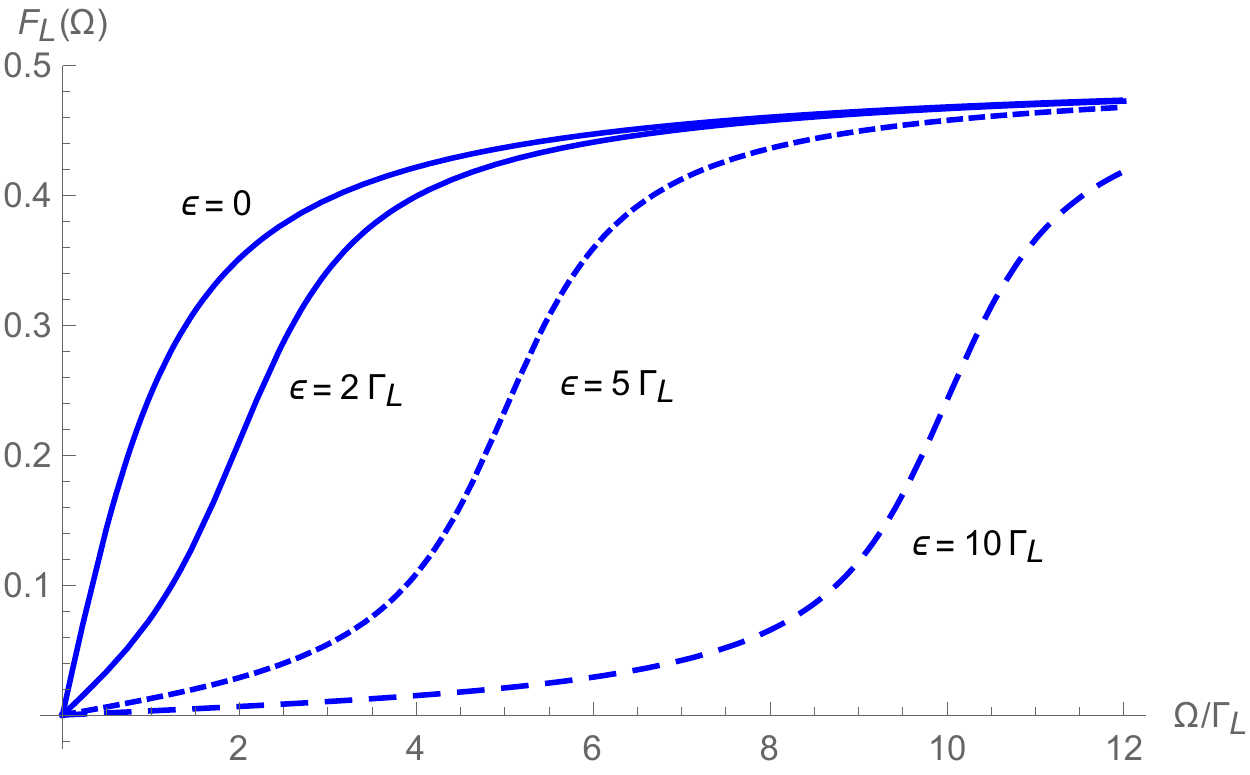}
\caption{(Color online.) The dimensionless function $F^{}_{L}(\Omega)$ [Eq. (\ref{FL})] for several values of $\epsilon$ measured with respect to the chemical potential on the lead, for $\Gamma^{}_{L}\beta^{}_{L}=10$, where $\beta^{}_{L}$ is the inverse temperature of the metal lead.}
\label{FO}
\end{figure}

 The matrix $W^{}_{L}(t,\omega)$ represents
 the correlation of the Aharonov-Casher phase factors at different times,
 \begin{align}
W^{}_{L}(t,\omega)&=\Big |\int^{t}_{}dt^{}_{1}e^{i(\omega-\epsilon+i\Gamma^{}_{L})(t-t^{}_{1})}V^{\dagger}_{L}(t^{}_{1})\Big |^{2}\ ,
\label{Wm}
\end{align}
 and is calculated in Appendix \ref{Technical}.
 The dc spin accumulation on the dot results from the corresponding  dc part of $W_{L}$ which involves the effective Zeeman interaction,  i.e., from the last term on the right hand-side of  Eq. (\ref{dcWL}),
  \begin{align}
 {\bf M}^{\rm dc}_{d}=2\hat{\bf x}\gamma\alpha^{2}_{L}%2i{\bf B}^{+}_{L}\times{\bf B}^{-}_{L }
 F^{}_{L}(\Omega)\ ,
 \label{Mdotdc}
 \end{align}
where $F_{L}(\Omega)$ is an odd  function of $\Omega$,
 \begin{align}
 F^{}_{L}(\Omega)=\Gamma^{}_{L}\int\frac{d\omega}{2\pi}f^{}_{L}(\omega)[|D(\omega+\Omega)|^{2}-|D(\omega-\Omega)|^{2}]\ ,
 \label{FL}
 \end{align}
with \cite{com3} $|D(\omega)|^{2}=|\omega-\epsilon+i\Gamma^{}_{L}|^{-2}$. This function is depicted in Fig. \ref{FO}; as seen, the integrand (for $\epsilon>0$) is dominated by the resonance  of $D(\omega+\Omega)$ since the Fermi function (at low temperatures)  is non-zero only for  the negative $\omega'$s.
In Eq. (\ref{Mdotdc})  we have used  Eqs. (\ref{BLt}) to obtain
 $2i{\bf B}^{-}_{L}\times{\bf B}^{+}_{L} =- \hat{\bf x}\alpha^{2}_{L}\gamma$.
  The magnetization accumulated on the dot is indeed along $\hat{\bf x}$, as implied by the heuristic argument above.
The
 probability to magnetize the dot is determined by  the polarization of the time-dependent electric field. For  a linearly polarized electric field ($\gamma=0$)
 the effective magnetic field for the absorption  process is parallel to that of the emission, ${\bf B}^{-}_{L}\parallel {\bf B}^{+}_{L}$,
 leading to a vanishing magnetic order.
 In contrast, for   circular or elliptic polarization ($\gamma\neq 0$) there appears a dc magnetization on the dot, which is linear in  $\gamma$.

% %%%%%%%%%%%%%%%%%%%%%%%%%%%%%%%%%%%%%%%%%%%%%%%%%%%%%%%%%%
%%%%%%%%%%%%%%%%%%%%%%%%%%%%%%%%%%%%%%%%%%%%%%%%%%%%%%%%%%%%
%%%%%%%%%%%%%%%%%%%%%%%%%%%%%%%%%%%%%%%%%%%%%%%%%%%%%%%%%%%%
%%%%%%%%%%%%%%%%%%%%%%%%%%%%%%%%%%%%%%%%%%%%%%%%%%%%%%%%%%%%

Evidently [see Eq. (\ref{Mdtr})],  the magnetic order built on the dot has also an ac component which oscillates with the frequencies $\Omega$ and $2\Omega$, see
Eq.  (\ref{acWL}).
This component gives the temporal variation of the spin polarization  on the dot. In the following we add to this component the rate by which the  magnetic order is established on the lead, thus examining the total time dependence  of the  spin population in the entire system.

The magnetization rate in the metal lead, $\dot{\bf M}^ {}_{L}(t)$, is defined as
\begin{align}
\dot{\bf M}^ {}_{L}(t)=\frac{d}{dt}\sum_{\bf k}\sum_{\sigma,\sigma'}\langle c^{\dagger}_{{\bf k}\sigma}(t)c^{}_{{\bf k}\sigma'}(t)\rangle
\sig^{}_{\sigma\sigma'}\ ,
\label{defdML}
\end{align}
where the time derivative and the quantum average are  with respect to the Hamiltonian (\ref{HOm}). This rate can be expressed in terms of lesser Green's functions,  $G^{<}_{Ld}$ and $G^{<}_{dL}$,  defined in Eqs. (\ref{GM}),
\begin{align}
\frac{d}{dt}\sum_{\bf k}\langle c^{\dagger}_{{\bf k}\sigma}(t)c^{}_{{\bf k}\sigma'}(t)\rangle&=J^{}_{L}
[G^{<}_{Ld}(t,t)V^{\dagger}_{L}(t)\nonumber\\
&-V^{}_{L}(t)G^{<}_{dL}(t,t)]^{}_{\sigma'\sigma}\ .
\label{rate1}
\end{align}
By solving the corresponding Dyson equations [Eqs. (\ref{DYGM})],
 one obtains  this magnetization rate, % on an arbitrary unit vector $\hat{\el}$ ($\hat{\el}=\hat{\bf x},\hat{\bf y},\hat{\bf z})$,
  \begin{align}
\dot{\bf M}^{}_{L}(t)={\rm Tr}
\{X^{}_{L}(t)V^{\dagger}_{L}(t)\sig V^{}_{L}(t)\}\ .
\label{dMLlm}
\end{align}
Here we have introduced the matrix
\begin{align}
X^{}_{L}(t)&=i\frac{d G^{<}_{dd,L}(t,t)}{dt}\nonumber\\
&=-2\Gamma^{}_{L}\int\frac{d\omega}{2\pi}f^{}_{L}(\omega)\frac{\partial W^{}_{L}(t,\omega)}{\partial t}\ .
\label{XLtm}
\end{align}
 [The derivation is contained in  Eqs. (\ref{X})-%, (\ref{dGd}), and
(\ref{XLt}).]
Comparing Eqs. (\ref{Mdtr}) and (\ref{dMLlm}),
we find
that while the (oscillating) rate of change of the magnetic moment on the dot is
\begin{align}
\dot{\bf M}^{}_{d}(t)%\cdot\hat{\el}
=-
{\rm Tr}
\{X^{}_{L}(t)\sig%\cdot\hat{\el}
 \}\ ,
\label{MLdot}
\end{align}
that in the lead, Eq. (\ref{dMLlm}), in addition to a sign difference,
requires a rotation of  $\sig$ by the Aharonov-Casher phase factors,
\begin{align}
\sig\rightarrow V^{\dagger}_{L}(t)\sig V^{}_{L}(t)\ .
\label{elpsi}
\end{align}
The total rate of the spin population in the junction  is
\begin{align}
&\dot{\bf M}^{}_{L}(t)+\dot{\bf M}^{}_{d}(t)={\rm Tr}\{X^{}_{L}(t)\big[V^{\dagger}_{L}(t)\sig V^{}_{L}(t)-\sig\big]\}
\ ,
\label{totdM}
\end{align}
%\mats{[MJ: Should it not be
%\begin{align}
%&\dot{\bf M}^{}_{L}(t)+\dot{\bf M}^{}_{d}(t)={\rm Tr}\{X^{}_{L}(t)\big[V^{\dagger}_{L}(t)\sig V^{}_{L}(t) - \sig\big]\}
%\ ?
%\label{totdM}
%\nonumber
%\end{align}
%}
and it vanishes only if there is no rotation, i.e., $V^{}_{L}(t)={\bf 1}$.
Put differently,
the total rate of the spin population along an arbitrary direction $\hat{\el}$, is
\begin{align}
&\hat{\el}\cdot[\dot{\bf M}^{}_{L}(t)+\dot{\bf M}^{}_{d}(t)]=2\Gamma^{}_{L}\int\frac{d\omega}{2\pi}f^{}_{L}(\omega)\nonumber\\
&\times{\rm Tr}\Big \{\frac{\partial W^{}_{L}(t,\omega)}{\partial t}\sig\cdot[\hat{\el
}-\hat{\el}'_{L}(t)
]\Big \}\ ,
\label{totdM}
\end{align}
where $\hat{\el}'_{L}(t)$ is the direction obtained upon rotating $\hat{\el}$ by the Aharonov-Casher phase factors,
\begin{align}
\sig\cdot\hat{\el}'_{L}(t)=V^{\dagger}_{L}(t)\sig\cdot\hat{\el} V^{}_{L}(t)\ .
\label{elp}
\end{align}
The deviation of $\hat{\el}'_{L}(t)$ away from $\hat{\el}$ %,  $\hat{\el}'(t)-\hat{\el}$,
 determines the amount by which the magnetization in the entire system is not conserved for a fixed direction, $\hat{\el}$.

\begin{figure}[htp]
\includegraphics[width=7cm]{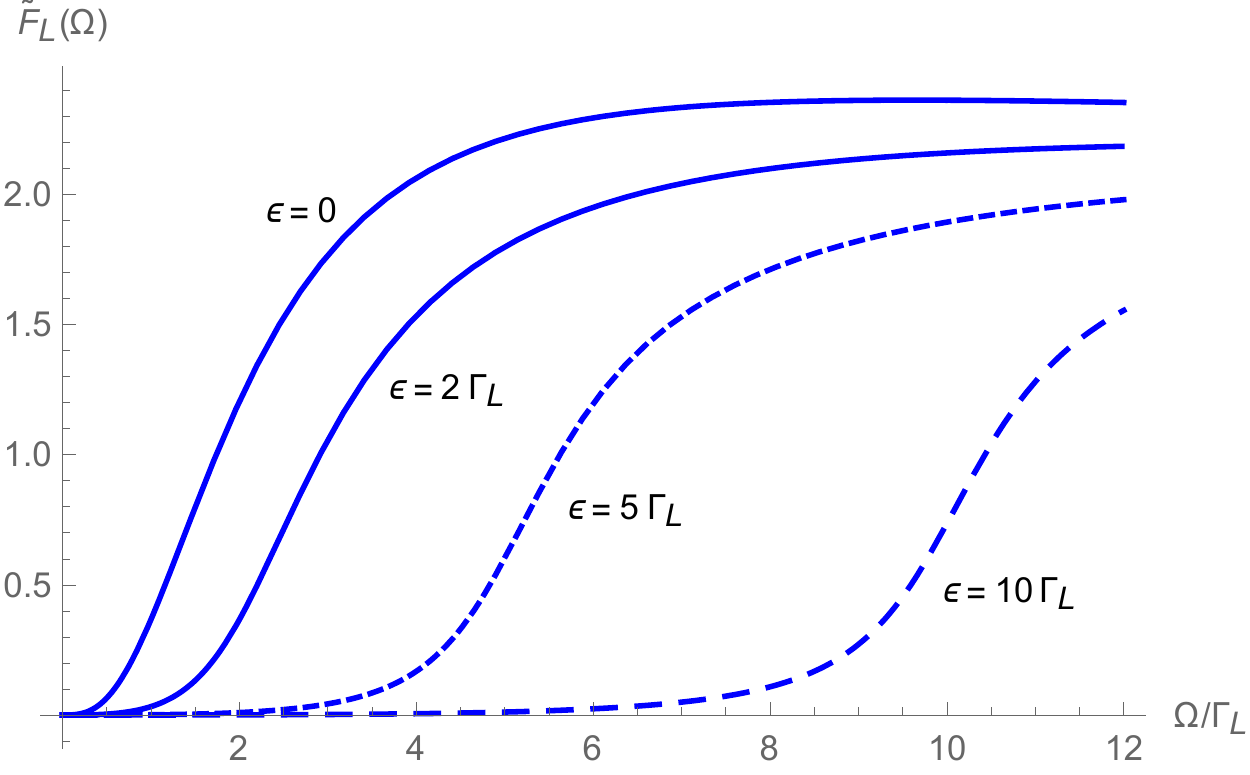}
\caption{(Color online.) The dimensionless function $\widetilde{F}^{}_{L}(\Omega)$ [Eq. (\ref{tildFL})] for several values of $\epsilon$ measured with respect to the chemical potential on the lead, for $\Gamma^{}_{L}\beta^{}_{L}=10$, where $\beta^{}_{L}$ is the inverse temperature of the metal lead.}
\label{tilFO}
\end{figure}

 Interestingly, the non-conservation has a dc component.
Up to second order in the spin-orbit coupling, it suffices  to consider the rotation
 to linear order in the spin-orbit coupling \cite{com2}
\begin{align}
\hat{\el'}^{}_{L}(t)\sim
\hat{\el}+2[{\bf B}^{+}_{L}e^{-i\Omega t}+{\bf B}^{-}_{L}e^{i\Omega t}]\times\hat{\el}
\label{elpa}\ .
\end{align}
Introducing this expression into Eq. (\ref{dMLlm})
%(\ref{totdM})}}
[and making use of Eqs. (\ref{elp}) and (\ref{acWL})],
one finds that the total rate in the junction includes two contributions: an oscillating part, which exists in both the lead and in the dot, and a dc part, which exists only in the lead (since the non-oscillating dot magnetization is constant in time),  along the $\hat{\bf x}$-axis,
\begin{align}
&\dot{\bf M}^{}_{L}(t)\Big |^{\rm dc}_{}=2 \hat{\bf x}\gamma\alpha^{2}_{L}\Gamma^{}_{L}
\widetilde{F}^{}_{L}(\Omega)\ ,
\label{dMLd}
\end{align}
where
\begin{align}
\widetilde{F}^{}_{L}(\Omega)&=4\Gamma^{}_{L}\Omega^{2}\int\frac{d\omega}{2\pi}f^{}_{L}(\omega)
|D(\omega)|^{2}\nonumber\\
&\times[|D(\omega+\Omega)|^{2}-|D(\omega-\Omega)|^{2}]\  .
\label{tildFL}
\end{align}
This function is plotted in Fig. \ref{tilFO}.
As seen, this dc component of the rate is along the $\hat{\bf x}$-axis,
just like the dc magnetization on the dot, Eq. (\ref{Mdotdc}), both quantities being odd in the microwave frequency,
$\Omega$.

% %%%%%%%%%%%%%%%%%%%%%%%%%%%%%%%%%%%%%%%%%%%%%%%%%%%%%%%%%%
%%%%%%%%%%%%%%%%%%%%%%%%%%%%%%%%%%%%%%%%%%%%%%%%%%%%%%%%%%%%
%%%%%%%%%%%%%%%%%%%%%%%%%%%%%%%%%%%%%%%%%%%%%%%%%%%%%%%%%%%%
%%%%%%%%%%%%%%%%%%%%%%%%%%%%%%%%%%%%%%%%%%%%%%%%%%%%%%%%%%%%

The total magnetization in the system along a fixed (in time) direction $\hat{\el}$ is not conserved. However,  one may examine possible cancellations of the magnetization rates.
Adding  the magnetization rate in the  dot along $\hat{\el}$, to that in the lead along a time-dependent vector given by
 $\hat{\el}''_{L}(t)$,
\begin{align}
\sig\cdot\hat{\el}''_{L}(t)=V^{}_{L}(t)\sig\cdot\hat{\el} V^{\dagger}_{L}(t)\ ,
\label{elph}
\end{align}
results in
\begin{align}
{\rm Tr}
\{X^{}_{L}(t)V^{\dagger}_{L}(t)\sig\cdot\hat{\el}''_{L}(t) V^{}_{L}(t)\}+\dot{\bf M}^{}_{d}(t)\cdot\hat{\el}=0
\ ,
\end{align}
%\mats{[MJ: Why not right the above equation as
%\begin{align}
%\dot{\bf M}^{}_{L}(t)\cdot\hat{\el}''_{L}(t)+\dot{\bf M}^{}_{d}(t)\cdot\hat{\el}=0
%\ ,
%\end{align}
which implies that the sum of the spin currents along these specific directions vanishes.
The sum of the dot magnetization along $\hat{\el}$ and of the lead magnetization along $\hat{\el}''_L(t)$ is conserved.
This is physically understood: an electron  magnetization  along $\hat{\el}$ in the dot rotates by the Aharonov-Casher factor to
be along $\hat{\el}''_L(t)$ in the lead.

% %%%%%%%%%%%%%%%%%%%%%%%%%%%%%%%%%%%%%%%%%%%%%%%%%%%%%%%%%%
%%%%%%%%%%%%%%%%%%%%%%%%%%%%%%%%%%%%%%%%%%%%%%%%%%%%%%%%%%%%

 \section{A dot coupled to two metal reservoirs}
 \label{2r}

\begin{figure}[htp]
\includegraphics[width=7cm]{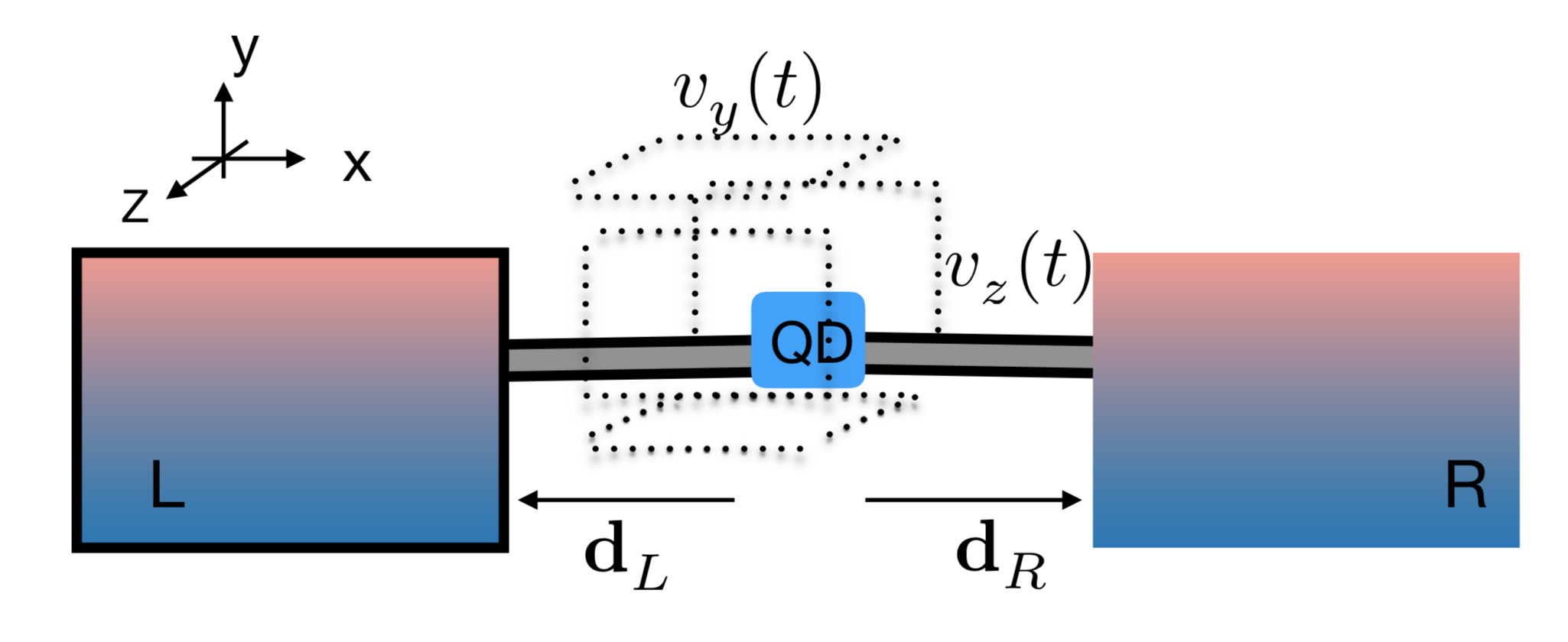}
\caption{(Color online.)  Illustration of a junction comprising a   quantum dot,  attached by two weak links lying along  the $\hat{\bf x}$-axis  to two reservoirs, denoted $L$ and $R$. As in Fig. \ref{single_lead},  the four plates mark the application of microwave-induced ac gate voltages, $v_{y}(t)$ and $v_{z}(t)$.  These give rise to time-dependent spin-orbit interactions in the weak links.}
\label{sys}
\end{figure}

The main reason for extending our scheme to a dot coupled to more than a single lead [see Fig. \ref{sys}], is to explore the possibility that  the induced spin-orbit interaction in, say, the left weak link, will generate a magnetic moment in the right lead. In other words, we wish to find out how the existence of one lead affects  the accumulated spin magnetization in the other.

Consider the magnetization rate in the left lead $\dot{\bf M}^{}_{L}(t)$, as defined in Eqs. (\ref{defdML}) and (\ref{rate1}),  when applied to the two-terminal junction depicted in Fig. \ref{sys}. % Its explicit calculation is detailed in Appendix \ref{Technical}.
It  is again convenient to express this quantity in terms of the (matrix) function $X^{}_{L}(t)$,
{\it cf.}  Eq. (\ref{dMLlm}).
However, in contrast to the configuration dealt with in Sec. \ref{AMs}, in the case where the dot is coupled to two leads, $X^{}_{L}(t)$ takes the form
\begin{align}
&X^{}_{L}(t)=2\int\frac{d\omega}{2\pi}\Big (-\Gamma^{}_{L}f^{}_{L}(\omega)\frac{\partial W^{}_{L}(t,\omega)}{\partial t}
\nonumber\\
&+2\Gamma^{}_{L}\Gamma^{}_{R}
[f^{}_{R}(\omega)W^{}_{R}(t,\omega)-f^{}_{L}(\omega)W^{}_{L}(t,\omega)]\Big )\ .
\label{XL2l}
\end{align}
[This expression results upon inserting Eq. (\ref{Gd2s}) for $G_{dd,L}$--and the corresponding one for $G^{}_{dd,R}$-- into Eq. (\ref{XLt}).]
The analogous function $X_{R}^{}(t)$ is obtained from Eq.
(\ref{XL2l})
by replacing  $L$ with $R$.

% %%%%%%%%%%%%%%%%%%%%%%%%%%%%%%%%%%%%%%%%%%%%%%%%%%%%%%%%%%
%%%%%%%%%%%%%%%%%%%%%%%%%%%%%%%%%%%%%%%%%%%%%%%%%%%%%%%%%%%%
%%%%%%%%%%%%%%%%%%%%%%%%%%%%%%%%%%%%%%%%%%%%%%%%%%%%%%%%%%%%

  The detailed calculation of the rate $\dot{\bf M}^{}_{L}(t)$ is carried out in Appendix \ref{Technical}, see Eq. (\ref{dMLA}) there. The dc component  is presented here,
 \ \begin{align}
& \dot{\bf M}^{\rm dc}_{L}=
\hat{\bf x}\Big (\gamma\alpha^{2}_{L}[2\Gamma\widetilde{F}^{}_{L}(\Omega)-4\Gamma^{}_{R}F^{}_{L}(\Omega)]%\Big (-i\Omega\Gamma^{}_{L}
%\int\frac{d\omega}{2\pi}f^{}_{L}(\omega)[
%D^{}_{3}(\omega)-D^{\ast}_{3}(\omega)]
\nonumber\\
&+\gamma\alpha^{2}_{R}%2i {\bf B}^{+}_{R}\times{\bf B}^{-}_{R}
4\Gamma^{}_{L}F^{}_{R}(\Omega)-\gamma\alpha^{}_{L}F^{}_{LR}(\Omega)\Big )\ ,
\label{dMLdc}
\end{align}
where $F^{}_{L}(\Omega)$ is defined in Eq. (\ref{FL}), $F_{R}(\Omega)$ is derived from the same equation by replacing $L$ with $R$, $\widetilde{F}^{}_{L}(\Omega)$ is defined in Eq. (\ref{tildFL}),
and
\begin{align}
F^{}_{LR}(\Omega)=&8\Gamma^{}_{L}\Gamma^{}_{R}
\int\frac{d\omega}{2\pi}[\alpha^{}_{R}f^{}_{R}(\omega)\nonumber\\
&+\alpha^{}_{L}f^{}_{L}(\omega)]2{\rm Re}[D^{}_{3}(\omega)]\ ,
\label{FLR}
\end{align}
with
\begin{align}
2{\rm Re}[D^{}_{3}(\omega)]&= 4|D(\omega-\Omega)D(\omega+\Omega)|^{2}\nonumber\\
&\times\Omega(\omega-\epsilon)
[1-\Omega^{2}|D(\omega)|^{2}] \ .
 \end{align}

Adding the rate of change of the magnetization in the  left lead [using   Eqs. (\ref{dMLlm}) and the analogous one for the right lead]  to the analogous one for the rate of change of the  magnetization in the right lead, $\dot{\bf M}^{}_{R}$,
yields
\begin{align}
[\dot{\bf M}^{}_{L}(t)+
\dot{\bf M}^{}_{R}(t)]\cdot\hat{\el}&={\rm Tr}\{X^{}_{L}(t)\hat{\el}'_{L
}(t)\cdot\sig\}\nonumber\\
&+X^{}_{R}(t)\hat{\el}'_{R}(t)\cdot\sig\}\ ,
\label{dotMLR}
\end{align}
where $\hat{\el}$ is again an arbitrary direction,  and $\hat{\el}'_{L}(t)$, defined in Eq. (\ref{elp}), is the direction reached upon rotating  $\hat{\el}$  by the (time-dependent) Aharonov-Casher factors of the left link. Similarly,
$\hat{\el}'_{R}(t)$ is the direction reached by the rotation with the Aharonov-Casher factors of the right link.
The rate of change of the magnetization in the dot, $\dot{\bf M}^{}_{d}(t)$, comprises contributions from the coupling with the left reservoir and the right one (see Appendix \ref{Technical}). The first is given in Eq.
 (\ref{Mdtr}), and the second is obtained from it by replacing $L$ with $R$. Thus, its rate of change is
\begin{align}
\dot {\bf M}^{}_{d}(t)\cdot\hat{\el}=2
\int&\frac{d\omega}{2\pi}{\rm Tr}\Big \{\Big (\Gamma^{}_{L}f^{}_{L}(\omega)\frac{dW^{}_{L}(t,\omega)}{dt}\nonumber\\
&+
 \Gamma^{}_{R}f^{}_{R}(\omega)\frac{dW^{}_{R}(t,\omega)}{dt}\Big )\sig\cdot\hat{\el}\Big \}
 \ .
 \label{fullMd}
 \end{align}
Adding together Eqs. (\ref{dotMLR}) and (\ref{fullMd}) [using Eq.
(\ref{XL2l})
and the analogous one for $X_{R}(t)$] gives the total rate of change of the magnetization in the two-terminal junction
along
an arbitrary direction $\hat{\el}$,
\begin{widetext}
\begin{align}
[\dot{\bf M}^{}_{d}(t)+\dot{\bf M}^{}_{L}(t)+\dot{\bf M}^{}_{R}(t)]\cdot\hat{\el}&=2{\rm Tr}\Big \{
\int\frac{d\omega}{2\pi}\Big (\Gamma^{}_{L}f^{}_{L}(\omega)\frac{dW^{}_{L}(t,\omega)}{dt}\sig\cdot[\hat{\el}-\hat{\el}'^{}_{L}(t)]+
 \Gamma^{}_{R}f^{}_{R}(\omega)\frac{dW^{}_{R}(t,\omega)}{dt}\sig\cdot[\hat{\el}-\hat{\el}'^{}_{R}(t)]\Big )\Big \}\nonumber\\
 &+4\Gamma^{}_{L}\Gamma^{}_{R}
 {\rm Tr}\Big \{
\int\frac{d\omega}{2\pi}[f^{}_{L}(\omega) W^{}_{L}(t,\omega)-f^{}_{R}(\omega)W^{}_{R}(t,\omega)] [\hat{\el}'^{}_{R}(t)-\hat{\el}'^{}_{L}(t)]\cdot\sig\Big \}\ .
 \end{align}
 \end{widetext}
% \mats{[MJ: Maybe on line 1 $2\Gamma_R$ should be $\Gamma_T$?]}
 As found in Sec. \ref{AMs}, the total magnetization would have been conserved had the rotations of the spin on their way between the dot and the leads been ignored.
 The amount by which the total magnetization is not conserved when measured  along a fixed (time-independent) direction $\hat{\el}$ is determined by the rotations of this direction  from the dot to the left lead and to the right one.
 Thus,  the time-dependent spin-orbit coupling generate a time-independent   magnetization, and the amount by which it is not conserved has also a dc part.

 %On the other hand, when combining the magnetization rate in the dot along $\hat{\el}$, that in the left lead along $\hat{\el}''_{L}(t)$ [Eq. (\ref{elph})], and the one in the right lead along
%$\sig\cdot\hat{\el}''_{R}(t)=V^{}_{R}(t)\sig\cdot\hat{\el} V^{\dagger}_{R}(t)$ yields zero, i.e., the sum of the spin currents along these directions vanish.

 \vspace{0.5cm}

 %%%%%%%%%%%%%%%%%%%%%%%%%%%%%%%%%%%%%%%%%%%%%%%%%%%%%%%%%%
 \section{Discussion}
 \label{dis}
%%%%%%%%%%%%%%%%%%%%%%%%%%%%%%%%%%%%%%%%%%%%%%%%%%%%%%%%%%%%
We propose that inelastic tunneling of electrons through a weak link, accompanied by spin flips generated  by a spin-orbit coupling caused by a rotating electric field, is capable of producing a net spin population in a nonmagnetic device; the field can be induced by microwave radiation as indicated in Fig. \ref{single_lead}. The origin of this effect is the correlation between emission and absorption of photons by tunneling electrons and specific spin flips (from spin down to spin up or from spin up to spin down). Our conjecture was verified in Sec. \ref{AMs} for a single-level quantum dot
coupled to a nonmagnetic reservoir of electrons, in the particular case when the dot energy level, $\epsilon$, is situated above the Fermi energy, $\epsilon_F$, of the reservoir and hence is unoccupied at zero temperature.
However,
one can easily convince oneself that the effect is the same if the dot level is situated below the  the Fermi energy, $\epsilon < \epsilon_F$, so that the dot level is doubly occupied at zero temperature.

We would like to remind the reader that our calculation is carried out in
the weak electron tunneling limit. This means that the
probability for double occupancy of an initially empty
dot (the case discussed above) due to inelastic tunneling is negligibly small. Therefore, the
intra-dot Coulomb repulsion energy $U$ in a doubly occupied dot  does not enter the calculation. By
the same argument, if the dot is initially doubly occupied only one electron can be removed due to inelastic
tunneling in the weak tunneling limit. In this case the Coulomb energy $U$ does play a role, 
 since the energy cost of removing one electron from the dot is $\epsilon^{}_F-(\epsilon + U)$ (compared to the energy $\epsilon-\epsilon^{}_F$ required to add one  electron to an empty dot). This allows one to
expect that if the dot contains several energy levels that can be involved in photon-assisted spin-flip transitions, the amount of spin accumulation on the dot can be augmented compared to when the dot has only one level. Except for this difference, the role of the interaction energy is trivial.

As discussed in Sec. \ref{AMs}, photon absorption processes dominate at low temperatures. For a circularly polarized electric field  rotating in, say, the clockwise direction (in the sense defined in Sec. \ref{AMs}) the requirement that spin angular momentum  is conserved then only allows spins to flip from ``down" to ``up". For an unoccupied dot, $\epsilon > \epsilon_F$, this means that only transitions from an occupied electron state with spin down in the reservoir to the spin-up state in the dot are allowed.
If, on the other hand $\epsilon < \epsilon_F$, only transitions from an occupied spin-down state in the dot to an unoccupied electron spin-up state in the reservoir are allowed, leaving an uncompensated spin-up electron on the dot.
Consequently, inelastic transitions between electron states in the lead and both occupied ($\epsilon<\epsilon_F$) and unoccupied ($\epsilon>\epsilon_F$) dot states result in the same spin state on the dot. This allows one to
expect that if the dot contains several energy levels that can be involved in photon-assisted spin-flip transitions, the amount of spin accumulation on the dot can be augmented compared to when the dot has only one level.

Driving the electron spin dynamics by a rotating electric field as suggested in this paper represents only one of several options for achieving a time-dependent spin-orbit coupling in nanodevices. Another possibility is to use a mechanical drive by temporally modulating the geometry of the device \cite{Shekhter}.  A related recent theoretical idea  \cite{Murakami} proposes to exploit externally excited  chiral phonon modes in graphene (which cause the carbon atoms to rotate and hence the spin-orbit interaction to be time dependent) to accumulate spin and generate magnetization.

The Keldysh Green's function for the dot, defined in Eq. (\ref{sd}), can be viewed as the spin density matrix of a spin q-bit. Its quantum coherent dynamics is fully determined by the time dependence of the spin-orbit interaction, which is induced by the ac gate voltages [see Eq. (\ref{dGd})]. Hence, driving the device by microwaves as envisaged here offers the possibility to create and manipulate a spin q-bit by applying appropriate microwave pulses as is well-known from the field of quantum computing.

The results presented in this paper open the possibility to use microwave radiation  to activate a magnetic pattern at the surface of a conductor. An array of quantum dots could be deposited on the surface, each dot individually coupled to the conductor by spin-orbit-active tunnel junctions. The magnetization of each dot could in principle be controlled locally by electrostatic gates  or by mechanical deformations of the tunneling weak links. In this way, one might be able to create a multiple q-bit structure in which communication between the dots would be governed by spin currents flowing between the dots and the common reservoir. A study of such possibilities is well beyond the scope of the present paper,
 but might serve as a motivation for  further investigations of the possibility to create static magnetization by irradiation with microwaves.

%%%%%%%%%%%%%%%%%%%%%%%%%%%%%%%%%%%%%%%%%%%%%%%

%%%%%%%%%%%%%%%%%%%%%%%%%%%%%%%%%%%%%%%%%%%%%%%%%%%%%%

\begin{acknowledgments}
This research was  partially supported by the Israel Science Foundation (ISF), by the infrastructure program of Israel Ministry of Science and Technology under contract 3-11173, and  by the Pazy Foundation. We acknowledge the hospitality of the PCS at IBS, Daejeon, Korea [where part of this work was  supported by
IBS funding number
(IBS-R024-D1)], and Zhejiang University, Hangzhou, China.
\end{acknowledgments}

%%%%%%%%%%%%%%%%%%%%%%%%%%%%%%%%%%%%%%%%%%%%%%%%%%%%%%
%%%%%%%%%%%%%%%%%%%%%%%%%%%%%%%%%%%%%%%%%%%%%%%%%%%%%%

\appendix
\section{Technical details}
\label{Technical}

\noindent{\it 1. The Green's functions in the time domain.}
For a dot coupled to two leads (Fig. \ref{sys}), the Hamiltonian (\ref {H0m}) is augmented by a term describing the right lead,
$\sum_{{\bf p} ,\sigma}\epsilon^{}_{ p}c^{\dagger}_{{\bf p}\sigma}c^{}_{{\bf p}\sigma}$. In addition,  the tunneling Hamiltonian in Eq. (\ref{HOm}) includes  a term yielding the  tunneling between the dot and the right lead
which takes the same form as in Eq. (\ref{HtunL}), with ${\bf k}$ replaced by ${\bf p}$ and  $L$ by $ R$. [Note that  ${\bf d}^{}_{R}=\hat{\bf x}d^{}_{R}$ and consequently ${\bf B}^{}_{R}(t)/\alpha^{}_{R}=-{\bf B}^{}_{L}(t)/\alpha^{}_{L}$.]

The Dyson equation for the Green's function on the dot, $G_{dd}(t,t')$ (in matrix notations in spin space), reads
\begin{align}
G^{}_{dd}(t,t')=g^{}_{d}(t,t')
&+\int dt^{}_{1}g^{}_{d}(t,t^{}_{1})[J^{}_{L}V^{\dagger}_{L}(t^{}_{1})G^{}_{Ld}(t^{}_{1},t')
\nonumber\\
&+J^{}_{R}V^{\dagger}_{R}(t^{}_{1})G^{}_{Rd}(t^{}_{1},t')]\ .
\label{DYG}
\end{align}
The first term in the square brackets results from the tunnel coupling with the left lead [see Eq. (\ref{HtunL})], and the second comes from the tunnel coupling with the right lead. The two other Green's functions  introduced in Eq. (\ref{DYG}) are
\begin{align}
G^{}_{dL}(t,t')&=\sum_{\bf k}G^{}_{d{\bf k}}(t,t')
\ ,\nonumber\\
G^{}_{Ld}(t,t')&=\sum_{\bf k}G^{}_{{\bf k}d}(t,t')
\ ,
\label{GM}
\end{align}
(with analogous definitions for $G^{}_{dR}$ and $G_{Rd}$).
The Dyson's equation (\ref{DYG}), as all other encountered below, refer to all three Keldysh Green's functions, the lesser (superscript $<$), the retarded (superscript $r$),  and the advanced (superscript $a$) \cite{Langreth,Jauho}.
In Eq. (\ref{DYG}),
$g_{d}(t,t')$ is the Green's function of the isolated dot; its retarded and  advanced forms are
\begin{align}
g^{r(a)}_{d}(t,t')&=\mp i\Theta(\pm t\mp t')\exp[-i\epsilon(t-t')]\ ,
\label{lesgd}
\end{align}
while the lesser function is zero since
 the isolated dot is assumed to be empty.

The Dyson's equations for the Green's functions (\ref{GM}) read
 (in matrix notations in spin space)
\begin{align}
G^{}_{Ld}(t,t')&=J^{}_{L}\int dt^{}_{1}g^{}_{L}(t,t^{}_{1})V^{}_{L}(t^{}_{1})G^{}_{dd}(t^{}_{1},t')\ ,\ \
\nonumber\\
G^{}_{dL}(t,t')&=J^{}_{L}\int dt^{}_{1}G^{}_{dd}(t,t^{}_{1})V^{\dagger}_{L}(t^{}_{1})g^{}_{L}(t^{}_{1},t')\ ,
\label{DYGM}
\end{align}
where $g_{L}(t,t')$ is Green's function of the decoupled left lead. Within the wide-band approximation \cite{Meir},   the   retarded, advanced, and lesser functions of the latter are
\begin{align}
g^{r(a)}_{L}(t,t')&=\mp i\pi{\cal N}^{}_{L}\delta (t-t')\ ,
\label{ragL}
\end{align}
and
\begin{align}
g^{<}_{L}(t,t')&=i\sum_{k}e^{-i\epsilon^{}_{k}(t-t')}f^{}_{L}(\epsilon^{}_{k})\nonumber\\
&=2\pi i{\cal N}^{}_{L}\int\frac{d\omega}{2\pi}e^{-i\omega(t-t')}f^{}_{L}(\omega)\ .
\label{lesgL}
\end{align}
The density of states of the left lead at the Fermi energy is denoted ${\cal N}^{}_{L}$, and $f^{}_{L}(\epsilon^{}_{k})$
is the Fermi function there.

The physical quantities  studied in the main text involve the lesser Green's functions at equal times.
Straightforward manipulations
of Eqs.  (\ref{DYG}) and (\ref{DYGM})  yield that $G^{<}_{dd}(t,t)$
comprises contributions from the coupling of the dot to the left and right leads,
\begin{align}
G^{<}_{dd}(t,t)=G^{<}_{dd,L}(t,t)+G^{<}_{dd,R}(t,t)\ ,
\label{Gds}
\end{align}
where
\begin{align}
G^{<}_{dd,L}(t,t)&=2i\Gamma^{}_{L}\int\frac{d\omega}{2\pi}f^{}_{L}(\omega)W^{}_{L}(t,\omega)\ .
\label{Gd2s}
\end{align}
(For more details, see Refs. \onlinecite{Odashima} and \onlinecite{PreSM}.)
Here,
$\Gamma^{}_{L}=2\pi J^{2}_{L}{\cal N}^{}_{L}$ is the partial width of the resonance on the dot, created by the tunnel coupling with the left  lead. An analogous expression pertains for $G^{<}_{dd,R}(t,t)$. The total resonance width on the dot is $\Gamma=\Gamma_{L}+\Gamma_{R}$.

The key player in our scheme is the 2$\times$2 matrix in spin space, $W_{L}(t,\omega)$, defined in Eq. (\ref{Wm}). Exploiting the expression for $V^{\dagger}_{L}(t)$ valid for a weak spin-orbit coupling [see Eq. (\ref{HtunL})],
we find
\begin{align}
&\int^{t}_{}dt^{}_{1}e^{i(\omega-\epsilon+i\Gamma)(t-t^{}_{1})}[1-|{\bf B}^{}_{L}(t^{}_{1})|^{2}/2-i\sig\cdot{\bf B}^{}_{L}(t^{}_{1})]\nonumber\\
&=D(\omega) [1-(1+\gamma^{2})\alpha^{2}_{L}/4
]-[(1-\gamma^{2})\alpha^{2}_{L}/4]F^{}_{2}(t,\omega)\nonumber\\
&-i\sig\cdot[{\bf B}^{-}_{L}e^{i\Omega t}D(\omega-\Omega)+{\bf B}^{+}_{L}e^{-i\Omega t}D(\omega+\Omega)]\ ,
\label{hW}
\end{align}
where $D(\omega)$ is \cite{com3}
\begin{align}
D(\omega)&=i/[\omega-\epsilon +i\Gamma]\ ,
\label{D}
\end{align}
and
\begin{align}
F^{}_{2}(t,\omega)&=\frac{1}{2}[e^{i2\Omega t}D(\omega-2\Omega)+e^{-i2\Omega t}D(\omega +2\Omega)]
\ .
\label{F2}
\end{align}

Using the result (\ref{hW}) in Eq. (\ref{Wm}), one finds that
\begin{align}
W^{}_{L}(t,\omega)=W^{\rm dc}_{L}(\omega)+W^{\rm ac}_{L}(t,\omega)
\ ,
\label{WLt}
\end{align}
where
$W^{\rm dc}_{L}(\omega)$ does not depend on time,
\begin{align}
W^{\rm dc}_{L}(\omega)=|\widetilde{D}(\omega)|^{2}+iD^{}_{1}(\omega)\sig\cdot{\bf B}^{-}_{L}\times{\bf B}^{+}_{L}\ ,
\label{dcWL}
\end{align}
and $W^{\rm ac}_{L}(t,\omega)$ oscillates with frequencies $\Omega$ and $2\Omega$,
\begin{align}
W^{\rm ac}_{L}(t,\omega)={\bf B}^{-}_{L}\cdot{\bf B}^{-}_{L}e^{2i\Omega t}D^{}_{2}(\omega)+{\rm c.c.}\nonumber\\
+i\sig\cdot[{\bf B}^{+}_{L}e^{-i\Omega t}D^{}_{3}(\omega)-{\rm c.c.}]\ .
\label{acWL}
\end{align}The  function $|\widetilde{D}(\omega)|^{2}$ in  Eq. (\ref{dcWL})
\begin{align}
|\widetilde{D}(\omega)|^{2}&=
|D(\omega)|^{2}-(1+\gamma^{2})(\alpha^{2}_{L}/2)\Big (|D(\omega)|^{2}\nonumber\\
&-[|D(\omega-\Omega)|^{2}+|D(\omega+\Omega)|^{2}]/2\Big )\ ,
\end{align}
is the correction (due to the spin-orbit coupling) of the Breit-Wigner resonance on the dot. The other functions in Eqs. (\ref{dcWL}) and (\ref{acWL}) are
\begin{align}
D^{}_{1}(\omega)&=
|D(\omega-\Omega)|^{2}-|D(\omega+\Omega)|^{2}\ , \nonumber\\
D^{}_{2}(\omega)&=[(\omega-\epsilon)^{2}_{}-(\Omega-i\Gamma)^{2}_{}]^{-1}_{}\nonumber\\
&-[1+4i\Gamma\Omega|D(\omega)|^{2}_{}][(\omega-\epsilon)^{2}_{}-(2\Omega-i\Gamma)^{2}_{}]^{-1}_{}\ ,\nonumber\\
D^{}_{3}(\omega)&=|D(\omega)|^{2}_{}[2\Omega(\omega-\epsilon)][(\omega-\epsilon)^{2}-(\Omega+i\Gamma)^{2}]^{-1}_{}
%\nonumber\\
%&-\frac{1+2i\Gamma\Omega|D(\omega)|^{2}_{}}{(\omega-\epsilon)^{2}_{}-(2\Omega-i\Gamma)^{2}_{}}
\ ,
\label{D123}
\end{align}
and they all vanish when $\Omega=0$.

%%%%%%%%%%%%%%%%%%%%%%%%%%%%%%%%%%%%%%%%%%%%%%%%%%%%%%
%%%%%%%%%%%%%%%%%%%%%%%%%%%%%%%%%%%%%%%%%%%%%%%%%%%%%%

\noindent{\it 2. The magnetization rates in the leads.}
By solving the Dyson's equations (\ref{DYGM}), the magnetization rate in the left lead,  given in Eqs. (\ref{defdML}) and (\ref{rate1}),  can be expressed in terms of the Green's functions on the dot \cite{PreSM},
\begin{align}
&\frac{d}{dt}\sum_{\bf k}\langle c^{\dagger}_{{\bf k}\sigma}(t)c^{}_{{\bf k}\sigma'}(t)\rangle=
-2i\Gamma^{}_{L}\Big ([V^{}_{L}(t)G^{<}_{dd}(t,t)V^{\dagger}_{L}(t)]^{}_{\sigma'\sigma}
\nonumber\\
&-\int\frac{d\omega}{2\pi}f^{}_{L}(\omega)\nonumber\\
&\times\int dt^{}_{1}[e^{-i\omega(t-t^{}_{1})}V^{}_{L}(t^{}_{1})G^{a}_{dd}(t^{}_{1},t)V^{\dagger}_{L}(t)
-{\rm H.c.}%e^{-i\omega(t^{}_{1}-t)}V^{}_{L}(t)G^{r}_{dd}(t,t^{}_{1})V^{\dagger}_{L}(t^{}_{1})
]^{}_{\sigma'\sigma}\Big )\ .
\end{align}
This expression is conveniently written in the form
\begin{align}
&\frac{d}{dt}\sum_{\bf k}\langle c^{\dagger}_{{\bf k}\sigma}(t)c^{}_{{\bf k}\sigma'}(t)\rangle=[V^{}_{L}(t)X^{}_{L}(t)V^{\dagger}_{L}(t)]^{}_{\sigma'\sigma}\ ,
\end{align}
where
\begin{align}
&X^{}_{L}(t)=-2i\Gamma^{}_{L}G^{<}_{dd}(t,t)+2i\Gamma^{}_{L}\int\frac{d\omega}{2\pi}f^{}_{L}(\omega)\nonumber\\
&\times\int dt^{}_{1}[e^{-i\omega(t-t^{}_{1})}V^{\dagger}_{L}(t)V^{}_{L}(t^{}_{1})G^{a}_{dd}(t^{}_{1},t)-{\rm H.c.}]%-2i\Gamma^{}_{L}G^{<}_{dd}(t,t)
\ .
\label{X}
\end{align}
The advantage of this representation is revealed when
Eqs. (\ref{Gd2s}) and (\ref{Wm}) are used to find
\begin{align}
&\frac{d}{dt}G^{<}_{dd,L}(t,t)=-2\Gamma G^{<}_{dd,L}(t,t)
+2\Gamma^{}_{L}\int\frac{d\omega}{2\pi}f^{}_{L}(\omega)\nonumber\\
&\times\int dt^{}_{1}[V^{\dagger}_{L}(t)e^{-i\omega(t-t^{}_{1})}V^{}_{L}(t^{}_{1})G^{a}_{dd}(t^{}_{1},t)-{\rm H.c.}]%+\{L\Rightarrow R\}\Big )
\ .
\label{dGd}
\end{align}
It then follows that
\begin{align}
X^{}_{L}(t)&=id G^{<}_{dd,L}(t,t)/dt\nonumber\\
&
+2i[\Gamma^{}_{R} G^{<}_{dd,L}(t,t)-\Gamma^{}_{L} G^{<}_{dd,R}(t,t)]\ .
\label{XLt}
\end{align}
For the single-lead junction,
considered in Sec. \ref{AMs},
$\Gamma^{}_{R}=0$, and therefore only the first term on the right hand-side of Eq. (\ref{XLt}) survives.
The corresponding expression for the two-terminal junction is obtained upon inserting
Eqs. (\ref{Gds}) and (\ref{Gd2s}) in Eq. (\ref{XLt}); this
yields Eq. (\ref{XL2l}) in the main text.

The explicit expression for the magnetization rate in the left lead is obtained by using Eq. (\ref{XL2l}) in Eq. (\ref{dMLlm}).
Denoting for brevity
%\begin{align}
$\hat{\el'}^{}_{L}(t)=V^{\dagger}_{L}(t)\hat{\el}V^{}_{L}(t)$,
%\end{align}
and using Eqs. (\ref{dcWL}) and  (\ref{acWL}),
we find
\begin{widetext}
\begin{align}
\dot{\bf M}^{}_{L}(t)\cdot\hat{\el}&=-4i\Gamma^{}_{L}\int\frac{d\omega}{2\pi}
f^{}_{L}(\omega)
\hat{\el'}^{}_{L}(t)\cdot\frac{d}{dt}
[{\bf B}^{+}_{L}e^{-i\Omega t}D^{}_{3}(\omega)-{\rm c.c.}]\nonumber\\
&+8i\Gamma^{}_{L}\Gamma^{}_{R}\int\frac{d\omega}{2\pi}D^{}_{1}(\omega)[{\bf B}^{-}_{R}\times{\bf B}^{+}_{R}f^{}_{R}(\omega)
-{\bf B}^{-}_{L}\times{\bf B}^{+}_{L}f^{}_{L}(\omega)]\cdot\hat{\el'}^{}_{L}(t)
\nonumber\\
&
-8i\Gamma^{}_{L}\Gamma^{}_{R}
\int\frac{d\omega}{2\pi}
\Big (f^{}_{R}(\omega)
[{\bf B}^{+}_{R}e^{-i\Omega t}D^{}_{3}(\omega)-{\rm c.c.}]
-
f^{}_{L}(\omega)
%\hat{\el'}^{}_{L}(t)\cdot
[{\bf B}^{+}_{L}e^{-i\Omega t}D^{}_{3}(\omega)-{\rm c.c.}]\Big )\cdot\hat{\el'}^{}_{L}(t)
\ ,
\label{dMLA}
\end{align}
where the functions $D_{1}^{}(\omega) $ and $D^{}_{3}(\omega)$ are defined in Eqs. (\ref{D123}).
 The dc magnetization rate (to second order in the spin-orbit coupling) is
 \begin{align}
\dot{\bf M}^{\rm dc}_{L}\cdot\hat{\el}&=-
8i\Gamma^{}_{L}\Omega{\bf B}^{+}_{L}\times{\bf B}^{-}_{L}\cdot\hat{\el}
\int\frac{d\omega}{2\pi}f^{}_{L}(\omega)2{\rm Im}[
D^{}_{3}(\omega)]%-D^{\ast}_{3}(\omega)]
\nonumber\\
&+8i\Gamma^{}_{L}\Gamma^{}_{R}\int\frac{d\omega}{2\pi}D^{}_{1}(\omega)[{\bf B}^{-}_{R}\times{\bf B}^{+}_{R}f^{}_{R}(\omega)
-{\bf B}^{-}_{L}\times{\bf B}^{+}_{L}f^{}_{L}(\omega)]\cdot\hat{\el}^{}
\nonumber\\
&+16
i\Gamma^{}_{L}\Gamma^{}_{R}\Big ({\bf B}^{+}_{R}\times{\bf B}^{-}_{L}\cdot\hat{\el}\int\frac{d\omega}{2\pi}
f^{}_{R}(\omega)D^{}_{3}(\omega)+{\bf B}^{+}_{L}\times{\bf B}^{-}_{R}\cdot\hat{\el}\int\frac{d\omega}{2\pi}
f^{}_{R}(\omega)D^{\ast}_{3}(\omega)\Big )
\nonumber\\
&-16i\Gamma^{}_{L}\Gamma^{}_{R}{\bf B}^{+}_{L}\times{\bf B}^{-}_{L}\cdot\hat{\el}\int\frac{d\omega}{2\pi}
f^{}_{L}(\omega)2{\rm Re}
[D^{}_{3}(\omega)]%+D^{\ast}_{3}(\omega)]
\ .
\end{align}

As ${\bf B}^{\pm}_{L}/\alpha^{}_{L}=-{\bf B}^{\pm}_{R}/\alpha^{}_{R}$, and by Eqs. (\ref{D123})
\begin{align}
&2{\rm Re}[D^{}_{3}(\omega)]%+D^{\ast}_{3}(\omega)
=4|D(\omega-\Omega)D(\omega+\Omega)|^{2}\Omega(\omega-\epsilon)
[1-\Omega^{2}|D(\omega)|^{2}]\ ,\nonumber\\
&2{\rm Im}[D^{}_{3}(\omega)]%-i[D^{}_{3}(\omega)-D^{\ast}_{3}(\omega)]
=2\Gamma\Omega|D(\omega)|^{2}[|D(\omega-\Omega)|^{2}-|D(\omega+\Omega)|^{2}]\ ,
\end{align}
the rate $\dot{\bf M}^{\rm dc}_{L}$
takes the form given in Eq.
(\ref{dMLdc}).

\end{widetext}

%%%%%%%%%%%%%%%%%%%%%%%%%%%%%%%%%%%%%%%%%%%%%%%%%%%%%%

%%%%%%%%%%%%%%%%%%%%%%%%%%%%%%%%%%%%%%%%%%%%%%%%%%%%%%


\begin{thebibliography}{99}
%%%%%%%%%%%%%%%%%%%%%%%%%%%%%%%%%%%%%%%%%%%%%%%%%%%%%%
%%%%%%%%%%%%%%%%%%%%%%%%%%%%%%%%%%%%%%%%%%%%%%%%%%%%%%
\bibitem{Sander}
D. Sander, S. O. Valenzuela, D. Makarov, C. H. Marrows, E. E. Fullerton, P. Fischer, J. McCord, P. Vavassori, S. Mangin, P.  Pirro,
B. Hillebrands, A. D. Kent, T. Jungwirth, O. Gutfleisch, C. G. Kim, and A. Berger,
{\it The 2017 Magnetism Roadmap},
J. Phys. D: Appl. Phys. {\bf 50}, 363001 (2017).
%%%%%%%%%%%%%%%%%%%%%%%%%%%%%%%%%%%%%%%%%%%%%%%%%%%%%
\bibitem{Kirby}
B. J. Kirby, L. Fallarino, P. Riego, B. B. Maranville, C. W. Miller, and A. Berger,
{\it Nanoscale magnetic localization in exchange strength modulated ferromagnets},
Phys. Rev. B {\bf 98}, 064404 (2018).
%%%%%%%%%%%%%%%%%%%%%%%%%%%%%%%%%%%%%%%%%%%%%%%%%%%%%
\bibitem{Miron}
See e.g., J. M. Miron, G. Gaudin, S, Auffret, B. Rodmacq, A, Schuhl, S. Pizzini, J. Vogel, and P. Gambardella,
{\it Current-driven spin torque induced by the Rashba effect in a ferromagnetic metal layer},
Nature Materials {\bf 9}, 230 (2010).
%%%%%%%%%%%%%%%%%%%%%%%%%%%%%%%%%%%%%%%%%%%%%%%%%%%%%%
\bibitem{Pile}
S. Pile, M. Buchner, V. Ney, T. Schaffers, K. Lenz, R. Narkowicz, J. Lindner, H. Ohldag, and A. Ney,
{\it
Direct imaging of the ac component of the pumped spin polarization with element specificity,}
arXiv:2005.08728v1 (2020).
%%%%%%%%%%%%%%%%%%%%%%%%%%%%%%%%%%%%%%%%%%%%%%%%%%%%%%
\bibitem{Araki}
Y. Araki , T. Misawa , and K. Nomura,
{\it
Dynamical spin-to-charge conversion on the edge of quantum spin Hall insulator},
Phys. Rev. Research {\bf 2}, 023195 (2020).
%%%%%%%%%%%%%%%%%%%%%%%%%%%%%%%%%%%%%%%%%%%%%%%%%%%%%%
\bibitem{Rashba}
E. I. Rashba,
{\it Properties of semiconductors with an extremum loop .1. Cyclotron and combinational resonance in a magnetic field perpendicular to the plane of the loop}, Fiz. Tverd. Tela (Leningrad) {\bf 2}, 1224 (1960) [Sov.
Phys. Solid State {\bf 2}, 1109 (1960)]; Y. A. Bychkov and E. I. Rashba, {\it Oscillatory effects and the magnetic susceptibility of carriers in inversion layers},
J. Phys. C {\bf 17}, 6039 (1984).
%%%%%%%%%%%%%%%%%%%%%%%%%%%%%%%%%%%%%%%%%%%%%%%%%%%%%%
\bibitem{Edelstein}
V. M. Edelstein,
{\it Spin polarization of conduction electrons induced by electric current in two-dimensional asymmetric
electron systems},
Solid State Commun. {\bf 73}, 233 (1990).
%%%%%%%%%%%%%%%%%%%%%%%%%%%%%%%%%%%%%%%%%%%%%%%%%%%%%%
\bibitem{Rojas}
J. C. Rojas S\'{a}nchez, L. Vila, G.
Desfonds, S. Gambarelli, J. P. Attann\'{e},
J. M. De Teresa, C. Mag\'{e}n, and A. Fert,
{\it Spin-to-charge conversion using Rashba coupling at interface between non-magnetic materials},
Nature Comm. {\bf 4}, 2944 (2013).
%%%%%%%%%%%%%%%%%%%%%%%%%%%%%%%%%%%%%%%%%%%%%%%%%%%%%%
\bibitem{Salemi}
L. Salemi, M. Berritta, A. K. Nandy, and P. M. Oppeneer,
{\it Orbitally dominated Rashba-Edelstein effect in noncentrosymmetric antiferromagnets},
Nature Comm. {\bf 10}, 1038 (2019).
%%%%%%%%%%%%%%%%%%%%%%%%%%%%%%%%%%%%%%%%%%%%%%%%%%%%%%
\bibitem{Puebla}
J. Puebla, F. Auvray, N. Yamaguchi, M. Xu, S. Zulkarnaen Bisri, Y.
Iwasa, F. Ishii, and Y. Otani,
{\it Photoinduced Rashba spin-to-charge conversion via interfacial unoccupied state},
Phys. Rev. Lett. {\bf 122}, 256401 (2019).
%%%%%%%%%%%%%%%%%%%%%%%%%%%%%%%%%%%%%%%%%%%%%%%%%%%%%%
\bibitem{Hernangomez}
D. Hernang\'{o}mez-P\'{e}rez, J.  D. Torres,  and A. L\'{o}pez,
{\it Photoinduced electronic and spin properties of quantum Hall systems with Rashba spin-orbit coupling},
arXiv:2005.05450v1 (2020).
%%%%%%%%%%%%%%%%%%%%%%%%%%%%%%%%%%%%%%%%%%%%%%%%%%%%%%
\bibitem{AC}
Y. Aharonov and A. Casher,
 {\it Topological quantum Effects for Neutral Particles},
Phys. Rev. Lett. {\bf 53}, 319 (1984).
%%%%%%%%%%%%%%%%%%%%%%%%%%%%%%%%%%%%%%%%%%%%%%%%%%%%%%
\bibitem{PRL2013}
R. I. Shekhter, O. Entin-Wohlman, and A. Aharony,
{\it
Suspended nanowires as mechanically-controlled Rashba spin-splitters},
Phys. Rev. Lett. {\bf 111}, 176602 (2013).
%%%%%%%%%%%%%%%%%%%%%%%%%%%%%%%%%%%%%%%%%%%%%%%%%%%%%%
\bibitem{Bardarson}
J. H. Bardarson,
{\it A proof of the Kramers degeneracy of transmission eigenvalues from antisymmetry of the scattering matrix},
J. Phys. A: Math. Theor. {\bf 41}, 405203 (2008).
%%%%%%%%%%%%%%%%%%%%%%%%%%%%%%%%%%%%%%%%%%%%%%%%%%%%%
 \bibitem{R2019}
  O. Entin-Wohlman, R. I. Shekhter, M. Jonson, and A. Aharony,
  {\it Photovoltaic effect generated by spin-orbit interactions},
Phys. Rev. B {\bf 101}, 121303(R) (2020).
%%%%%%%%%%%%%%%%%%%%%%%%%%%%%%%%%%%%%%%%%%%%%%%%%%%%%%
\bibitem{Duckheim}
M. Duckheim and D. Loss,
{\it Electric Dipole Induced Spin Resonance in Disordered Semiconductors,}
Nature Phys. {\bf 2}, 195 (2006).

%%%%%%%%%%%%%%%%%%%%
\bibitem{com0} Although we treat the electric field classically,  a time-dependent quantum-mechanical perturbation theory with the Hamiltonian (\ref{HtunL}) will generate inelastic transitions to electron states with flipped spins and with  energy shifted by $\pm \Omega$ only. We can therefore refer to these energy quanta as `photons'.
     %%%%%%%%%%%%%%%%%%%%%%%%%%%%%%%%%%%%%%%%%%%%%%%%%%%%%%
\bibitem{com1}
The calculations in Appendix \ref{Technical} are carried out for a dot coupled to {\em two} leads, see Fig. \ref{sys}.
However, it is straightforward to infer from the expressions in Appendix \ref{Technical} the relevant quantities for the simpler junction depicted in Fig. \ref{single_lead}: all that one needs to do is to set $\Gamma^{}_{R}$, the partial resonance width on the dot resulting from the coupling to the right lead, to zero.
%%%%%%%%%%%%%%%%%%%%%%%%%%%%%%%%%%%%%%%%%%%%%%%%%%%%%%
\bibitem{com3}
The resonance width in $|D(\omega)|^{2}$ is $\Gamma_{L}$ when the dot is coupled to a single lead. For the two-terminal junction, the resonance width is $\Gamma=\Gamma^{}_{L}+\Gamma^{}_{R}$, see Appendix \ref{Technical}.
%%%%%%%%%%%%%%%%%%%%%%%%%%%%%%%%%%%%%%%%%%%%%%%%%%%%%%
\bibitem{com2}
It is sufficient to consider the rotation transformation to  first order in the spin-orbit coupling, since the other terms in Eq. (\ref{MLdot}) that contribute to the trace are at least first order in $\alpha^{}_{L}$.
%%%%%%%%%%%%%%%%%%%%%%%%%%%%%%%%%%%%%%%%%%%%%%%%%%%%%%
\bibitem{Shekhter}
R. I. Shekhter, O. Entin-Wohlman, M. Jonson, and A. Aharony,
{\it
Rashba spin-splitting of single electrons and Cooper pairs},
Low Temp. Phys. {\bf 43}, 303 (2017); {\it Photo-spintronics of spin-orbit active electric weak links},  {\it ibid.} {\bf 43}, 910 (2017).
%%%%%%%%%%%%%%%%%%%%%%%%%%%%%%%%%%%%%%%%%%%%%%%%%%%%%%
\bibitem{Murakami}
M. Hamada and S. Murakami,
{\it Conversion between electron spin and microscopic atomic rotation},
Phys. Rev. Research {\bf 2}, 023275 (2020).
%%%%%%%%%%%%%%%%%%%%%%%%%%%%%%%%%%%%%%%%%%%%%%%%%%%%%%
%\bibitem{SM}
% See Supplemental Material at {\url{http://link.aps.org/supplemental/10.1103/PhysRevB.101.121303}} for details of the Keldysh Green's functions' calculation.
%%%%%%%%%%%%%%%%%%%%%%%%%%%%%%%%%%%%%%%%%%%%%%%%%%%%%%
%%%%%%%%%%%%%%%%%%%%%%%%%%%%%%%%%%%%%%%%%%%%%%%%%%%%%%
\bibitem{Langreth}
D. C. Langreth, {\it Linear and nonlinear response theory with applications}, in Linear and Nonlinear Electron Transport in Solids, eds. J. T. Devreese and E. van Boren (Plenum, New York, 1976).
%%%%%%%%%%%%%%%%%%%%%%%%%%%%%%%%%%%%%%%%%%%%%%%%%%%%%%
\bibitem{Jauho}
A. P. Jauho, {\it Nonequilibrium Green function modelling of transport in mesoscopic systems}, in Progress in Nonequilibrium Green's Functions II, eds. M. Bonitz and D. Semkat (World Scientific, Singapore, 2003).
%%%%%%%%%%%%%%%%%%%%%%%%%%%%%%%%%%%%%%%%%%%%%%%%%%%%%%
 \bibitem{Meir}
A-P. Jauho, N. S. Wingreen, and Y. Meir,
{\it Time-dependent transport in interacting and noninteracting resonant-tunneling systems},
Phys. Rev. B {\bf 50}, 5528 (1994).
%%%%%%%%%%%%%%%%%%%%%%%%%%%%%%%%%%%%%%%%%%%%%%%%%%%%%%
\bibitem{Odashima}
M. M. Odashima and C. H. Lewenkopf,
{\it Time-dependent resonant tunneling transport: Keldysh and Kadanoff-Baym nonequilibrium Green's functions in an analytically soluble problem},
Phys. Rev. B {\bf 95}, 104301 (2017).
%%%%%%%%%%%%%%%%%%%%%%%%%%%%%%%%%%%%%%%%%%%%%%%%%%%%%%
\bibitem{PreSM}
 See Supplemental Material at {\url{http://link.aps.org/supplemental/10.1103/PhysRevB.101.121303}} for details of the Keldysh Green's functions' calculation.
%%%%%%%%%%%%%%%%%%%%%%%%%%%%%%%%%%%%%%%%%%%%%%%%%%%%%%









%%%%%%%%%%%%%%%%%%%%%%%%%%%%%%%%%%%%%%%%%%%%%%%%%%%%%%
\end{thebibliography}
 \end{document}